*Review*

# A Review of Electronic Transport in Superconducting Sr$_2$RuO$_4$ Junctions

Muhammad Shahbaz Anwar * and Jason W. A. Robinson *

Department of Materials Science & Metallurgy, 27 Charles Babbage Rd, Cambridge CB3 0FS, UK
* Correspondence: msa60@cam.ac.uk (M.S.A.); jjr33@cam.ac.uk (J.W.A.R.)

**Abstract:** We review electronic transport in superconducting junctions with Sr$_2$RuO$_4$. Transport measurements provide evidence for chiral domain walls and, therefore, chiral superconductivity in superconducting Sr$_2$RuO$_4$, but so far, the symmetry of the underlying superconducting state remains inconclusive. Further studies involving density of states measurements and spin-polarised transport in local/non–local Sr$_2$RuO$_4$ junctions with magnetic materials could lead to fundamental discoveries and a better understanding of the superconducting state.

**Keywords:** unconventional superconductivity; Sr$_2$RuO$_4$; chirality; chiral domain wall; superconducting junctions

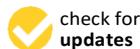

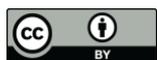

## 1. Introduction

Macroscopic quantum phenomenon of conventional superconductivity occurs with the formation of Cooper pairs of electrons in a spin–singlet state with opposite electronspins ($S = 0$) and a zero centre–of–mass moment [1]. Such singlet Cooper pairs exhibit even parity *s*-wave symmetry of the order parameter. Unconventional high–temperature superconductors (HTSs) are spin-singlet but have even parity and *d*-wave symmetry of the order parameter [2].

Spin triplet superconductivity with spin-polarised ($S = 1$) electron pairs (spin–triplet Cooper pairs) can emerge with a *p*–wave symmetry of the order parameter. Such *p*-wave superconductivity offers spin and orbital degrees of freedom. We note that spin-triplet Cooper pairs with odd-parity *s*–wave symmetry can be generated at an interface between an *s*-wave superconductor (SC) and a ferromagnet (F) [3,4]—for example, spin-mixing due to the ferromagnetic exchange field generates $m = 0$ spin-triplet Cooper pairs, and broken spin–rotational symmetry converts $m = 0$ triplets into equal spin $m = \pm 1$ triplet pairs [5–14]. Moreover, broken inversion symmetry at an SC/F interface generates spin–orbit coupling (SOC), which creates interfacial spin–triplet superconductivity [15–17]. Junctions based on spin–triplet correlations have the potential for quantum information technologies and superconducting spintronics [4].

There are a number candidate materials with possible *p*–wave spin–triplet superconductivity, including UPt$_3$, UTe$_2$ [18] and Sr$_2$RuO$_4$ (SRO214) [19]. The unconventional superconductor SRO214 has been intensively investigated to understand the symmetry of its underlying order parameter [20]. Recently, it has been shown that SRO214 may have a chiral *d*–wave spin–singlet instead of a chiral *p*–wave spin–triplet [21]. In this topical review, we focus on electrical transport properties of SRO214 superconducting junctions that may indicate the presence of a superconducting chiral domain.

## 2. Strontium Ruthenate: Sr$_2$RuO$_4$

SRO214 is a layered perovskite oxide superconductor with a K$_2$NiF$_4$–(nondistorted tetragonal with nonmagnetic ground state) crystal structure with space group *I*4/*mmm*, and it is isostructural with the parent compound La$_2$CuO$_4$ of an HTS, such as La$_{2-x}$Sr$_x$CuO$_6$ (a $d$-wave spin–singlet superconductor) [22]. The superconducting transition temperature ($T_c$) of SRO214 is low (1.5 K) compared with HTS materials.

Since the discovery of superconductivity in SRO214 [19], extensive experimental and theoretical works have been performed to investigate the symmetry of the superconducting order parameter [20–24]. Mackenzie et al. [25] investigated the effect of nonmagnetic impurities and found that the superconductivity is quenched as the residual resistivity increases to 1 μΩ·cm with an electron mean free path ($l_e$) below the in–plane superconducting coherence length $\xi$ = 66 nm. Furthermore, Mao et al. [26] observed a suppression of superconductivity in SRO214 single crystals with increasing crystal defects, implying that electron scattering suppresses $T_c$.

Internal magnetic fields in the superconducting state were measured in Kerr-rotation [27] and muon spin relaxation ($\mu$SR) [28,29], indicating that superconductivity in SRO214 breaks time reversal symmetry (TRS). This supports the chiral $p$–wave spin–triplet superconducting order parameter. In addition, electrical transport measurements of SRO214 junctions support dynamical behaviour that may attribute to broken TRS [30–36].

Missing edge currents [37] in the superconducting state, thermal conductivity indicating line nodes [38] and first order transition in $H_{c2}$ measured by heat capacity [39] do not support a chiral $p$–wave spin triplet. Recently, it was observed that both $T_c$ and $H_{c2}$ are enhanced with uniaxial pressure applied on SRO214 single crystals, which cannot be explained by the spin–triplet nature of the order parameter [40,41].

The spin degrees–of–freedom of the superconducting order parameter of SRO214 have been investigated by measuring the electron spin susceptibility. In earlier experiments, it was realised that the spin susceptibility of SRO214 is unchanged below $T_c$ [42,43]. For a spin–singlet superconductor, spin susceptibility is strongly suppressed, while it remains unchanged for a spin–triplet superconductor. In 1998, Ishida et al. [42] measured the Knight shift of an SRO214 single crystal using nuclear magnetic resonance (NMR) spectroscopy of $^{17}$O by applying a radio frequency (RF) pulse of 3.8 MHz and an in–plane magnetic field (parallel to RuO$_2$ planes). They found that the Knight shift as a function of temperature remains unchanged by crossing the temperature below $T_c$. It was also observed by polarised neutron scattering (PNS) experiments [43]. These findings were supporting the spin–triplet scenario with possible symmetry of the order parameter of $d = z(k_x \pm ik_y)$. However, Pustogow et al. [44] recently reported a significant decrease in the O$^{17}$ NMR Knight shift of SRO214 as a function RF-pulse energy at 20 mK. They found that in the previous NMR experiments the pulse energy ($\approx$ 80 μJ) was heating the SRO214 samples up to $T > T_c$. Similar results have been reproduced by other groups [45,46]. These observations show that the nature of the superconducting order parameter of SRO214 may be a spin–singlet, such as $s$- or $d$-wave. Recently, Grinenko et al. [47] performed the measurement of zero–field $\mu$SR on SRO214 single crystals as a function of temperature and uniaxial pressure. They observed stress–induced splitting between the $T_{c\text{-onset}}$ and broken TRS that relates with the chiral superconducting order parameter. Benhabib et al. [48] observed a sharp change in the shear elastic constant when crossing $T_c$, using the measurements of ultrasound velocity in SRO214. This indicates that the order parameter may exhibit a two–component nature.

In addition to experiments on SRO214 single crystals [20], various superconducting junctions based on SRO214 single crystals have been studied. The transport properties of these junctions show anomalies that may indicate unconventionality and dynamicity of the order parameter attributing to two-fold degenerate chiral domains formed because of broken TRS. In this review, we mainly focused on the transport properties of SRO214-based superconducting junctions.

A number of review articles have already been published that mainly focus on studies of SRO214 single–crystals [20–24]. This review focuses on phase-sensitive electrical properties of SRO214–based superconducting junctions to understand the superconducting order parameter. After a brief introduction, we discuss the transport properties of various superconducting junctions, including proximity structures, tunnel junctions, SQUIDs and Josephson effects. At the end, we provide an overview of candidate materials for spin-triplet superconducting junctions and summarise the article.

*Multiple Phases*

SRO214 belongs to the Ruddlesden–Popper Phase (RPP) family of Sr–based ruthenates $Sr_{n+1}Ru_nO_{3n+1}$, where $n$ is the number of $RuO_2$ layers in a unit cell. It is challenging to obtain pure SRO214 single crystals as impurity phases form as embedded inclusions during growth, including ferromagnetic $SrRuO_3$ (SRO113, $n = \infty$; $T_{curie}$ of 160 K), metamagnetic $Sr_3Ru_2O_7$ (SRO327, $n = 2$; $T_{curie} \approx 100$ K under 1 T magnetic field) and pure Ru metal. The floating zone technique results in high-quality SRO214 single crystals [49,50].

Multiple superconducting transitions are observed in SRO214 eutectic crystals with a broad $T_c$, extending up to 3 K and a sharp bulk transition with maximum $T_c$ of 1.5 K [51], depending on the quality of the crystal (Figure ??a). The 3–K–phase results from superconductivity that emerges at the interface between SRO214 bulk and inclusions of Ru or SRO327 in a eutectic crystal [52], possibly due to crystallographic strain at the interface [40,41].

The $H_{c2}$ of the 3–K–phase for both in- and out-of-plane directions are significantly higher than that for 1.5–K–phase, as shown in Figure 1b. It may suggest that both phases have different superconducting order parameters. Theoretically, the main difference between 1.5–K– and 3–K-phases is just the topology of the order parameter rather than the parity [53,54].

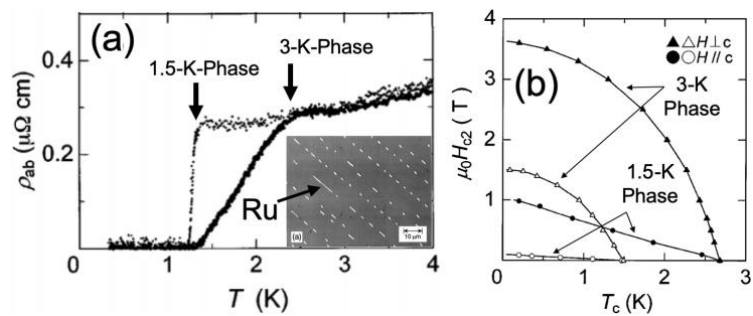

**Figure 1.** (**a**) Temperature–dependent–resistance $R(T)$ of pure (grey line) and eutectic (black line) SRO214 single crystals. The eutectic crystal exhibits multiple phases: the 3–K– and 1.5–K–phase. (**b**) Lower ($H_{c1}$) and upper critical fields ($H_{c2}$) vs. temperature of the 3–K– and 1.5–K–phase. (Adopted from [51] copyright APS 1998)

## 3. Half Quantum Fluxoid

In an SC ring, integer quantum fluxoid $\varphi = n\varphi_o$ (where $n$ is the integer and $\varphi_o$ is the flux quantum) stabilises because of the single–valuedness of the wave function $\psi_S = \Delta_o e^{i\theta}$, where $\Delta_o$ is the superconducting gap. The wave function of a spin–triplet equal–spin pairing superconductor $\psi_T = \Delta_o[\ e^{i\theta}\ |\uparrow\uparrow> + e^{i\theta}\ |\downarrow\downarrow>]$ has two degrees of freedom, which stabilise the half quantum fluxoid $\varphi = (n + 1/2)\varphi_o$.

To satisfy the magnetic flux quantisation, the applied magnetic field induces supercurrents in an SC ring. In the result, the free energy and $T_c$ of the superconductor oscillate, which leads to Little Parks (LP) oscillations [55]. The change in the $T_c$ can be investigated by measuring the magnetoresistance (MR) of an SC ring, particularly in the transition regime. A peak in MR oscillations of the ring corresponds to the switching in the supercurrent and free energy with induction of additional fluxoid, meaning a peak is the border between two integer quantum fluxoids. Therefore, peak splitting can occur with the stabilisation of an HQF.

Various research groups have investigated micro–sized rings of SRO214 and reported HQF behaviour. Jang et al. [56] observed half magnetisation steps in SRO214 rings ((outer radius ($r_{out}$) = 1.5–1.8 µm; internal diameter $r_{in} \approx 1$ µm; thickness ($t$) = 0.35 µm)) using cantilever magnetometry. Note that these steps may relate to the stabilisation of HQF as the steps are realised with an in–plane magnetic field applied in addition to out-of–plane magnetic fields. An in–plane magnetic field may be required to reduce the spin–current energy to stabilise the HQF [57]. Such HQFs are not observed in SRO214 rings with dimensions larger than the penetration depth $\lambda$. However, complicated fractional steps are induced with in–plane fields in a conventional SC NbSe$_2$ that are not half–steps and also depend on the direction of in–plane magnetic fields.

Measurements of MR may provide direct evidence of the existence of HQF in SRO214 rings. For this, Cai et al. [58,59] investigated MR as a function of magnetic fields applied along the *c*-axis (out-of–plane) of a circular SRO214 ring ($r_{out}$ = 1 µm; $r_{in}$ ≈ 0.5 µm; *t* = 0.2–0.6 µm). They observed that the amplitude of the MR oscillations is much higher than expected for LP's oscillations [55] and also observed well below the transition regime ($T < T_c$) and without applying an in–plane magnetic field. The peak splittings as a signature of HQF were not observed. That means that the observed MR behaviour may imitate the $I_c$ oscillations of a SQUID instead of LP's oscillations. The rings were fabricated using a focused ion beam (FIB) without any intentional weak links. However, fabrication related defects may develop weak links. Note that, in these experiments, an in-plane magnetic field was not applied, which is crucial to stabilise the HQF by minimising the spin–current energy [57].

Yasui et al. [60] investigated the MR of SRO214 rectangular rings with $r_{out}$ = 1.5 µm, $r_{in}$ ≈ 0.5 µm, *t* = 1.5 µm. Interestingly, the amplitude of MR oscillations with an out- of-plane magnetic field matches the expected amplitude of LP's oscillations (Figure 2). Moreover, weak signatures of peak splitting are detected even for no in–plane magnetic field. The splitting widened with an increase in-plane fields, which may correspond to the stabilisation of HQF.

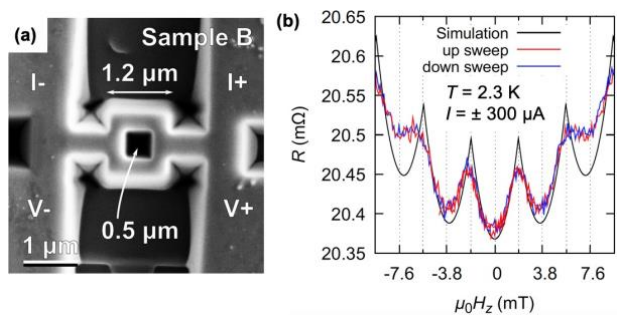

**Figure 2.** (**a**) Scanning electron micrograph of a rectangular SRO214 micro ring. (**b**) Magnetoresistance at 2.3 K with an excitation current of ±300 µA. The amplitude of oscillations matches well with the calculated Little–Parks oscillations. (Adopted from [60] copyright APS 2017)

This indicates that the electrical transport properties of an SRO214 ring may depend on the ring geometry. To understand this, recently, Yasui et al. [61] investigated SRO214 rings fabricated in two different geometries, as shown in Figure 3a,b. They observed SQUID-like oscillations in $I_c$ [58,59] at various different temperatures below $T_c$, as shown in Figure 3c. This indicates that these rings may consist of weak links. However, weak links were not intentionally prepared in the rings, implying that the domain wall may behave like a weak link creating Josephson junctions. They also performed theoretical calculations based on time-dependent Ginzburg–Landau simulations and considered chiral *p*–wave symmetry of the order parameter. It is realised that a ring can host a single-domain or multi–domain structure depending on the ratio of inner and outer radii ($r_{out}/r_{in}$). A multi–domain ring with restricted dimensions provides strong pinning to keep the domain walls within the arms of the ring. These domain walls provide weak links as Josephson coupling between two chiral domains. A phase diagram is shown in Figure 3e along with different ring geometries where colour maps indicate the density of the order parameter/ $|\psi|^2$. Kashiwaya et al. [32] studied micro rings with two weak links on each side, which may also provide pinning sites for domain walls. However, these rings do not show SQUID behaviour, only switching in $I_c$, which may relate to instabilities or dynamical behaviour of the superconducting order parameter.

Very recently Liu's group measured MR of SRO214 circular rings in the influence of in-plane magnetic field [62]. They observed splitting that may attribute to HQF, which supports the spin-triplet nature of superconductivity in SRO214.

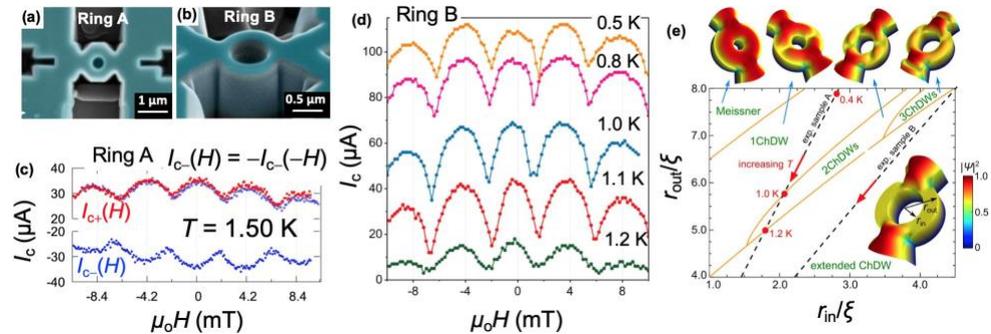

**Figure 3.** False-colour scanning electron micrographs of two SRO214 micro rings both have the same outer radius of $r_{out}$ = 0.55 µm, while (**a**) for ring A, the inner radius $r_{in}$ is 0.3 µm and (**b**) for ring B $r_{in}$ = 0.23 µm. (**c**) Magnetic–field–dependent critical current $I_c(H)$ at 1.50 K of ring A. (**d**) $I_c(H)$ of ring B measured at various temperatures. (**e**) Calculated geometry–dependent phase diagram with illustrations showing the various states within a ring at different points in the phase diagram. The colour maps show the Cooper–pair density $|\psi|^2$. In the absence of a magnetic field, for a wider ring, the Meissner state is stable, and the chiral domain walls are stabilised by reducing the width of the ring wall. (Adopted from [61] copyright NPG2020)

## 4. Broken Time Reversal Symmetry

A superconductor breaks TRS with the appearance of spontaneous magnetic fields below $T_c$. A conventional superconductor has perfect diamagnetism, which keeps TRS invariant; however, unconventional superconductors may break TRS. Therefore, measurements of spontaneous magnetic fields in the superconducting state provide the smoking gun proof for TRS.

It has been experimentally demonstrated that TRS is broken in SRO214, e.g., an increasing muon spin rotation ($\mu$SR) relaxation rate [28,47] and a nonzero Kerr rotation [29] below $T_c$ of SRO214 indicate the existence of an internal magnetic field. Based on the recent observation of transition temperatures splitting between $T_{c\text{-onset}}$ and TRS breaking, a temperature vs. stresses phase diagram for superconductivity in SRO214 has been presented (see Figure **??**) [47]. These observations suggest that SRO214 is a chiral superconductor and exhibits two–fold degenerate chiral domains with an opposite orbital angular moment separated by chiral–domain walls [63,64].

A direct observation of chiral–domains has not been reported [65]. For a chiral superconductor, edge currents at the sample edge or at domain walls are expected that may generate an out–of–plane magnetisation [66]. However, micro–scanning SQUIDs [37,67] and scanning Hall probes [68] have not detected the edge currents. This may indicate that edge currents may be balanced with Meissner screening currents or that SRO214 is a helical $p$–wave spin–triplet superconductor [63,64,67–70] or a spin-singlet superconductor.

There are five possible spin–triplet states compatible with a tetragonal crystal structure of SRO214, $E_u$ (Mulliken notation) a chiral $p$-wave state for which the $d$-vector is aligned along the $c$–axis and remaining four states ($A_{1u}$, $A_{2u}$, $B_{1u}$, and $B_{2u}$) are helical $p$–wave states with $d$–vectors that lie in the $ab$-plane [71]. The TRS is a qualitative difference between a chiral and four helical states; for chiral the $p$–wave state, TRS is broken and nonzero edge currents must appear, but for helical states, TRS is invariant with zero edge currents.

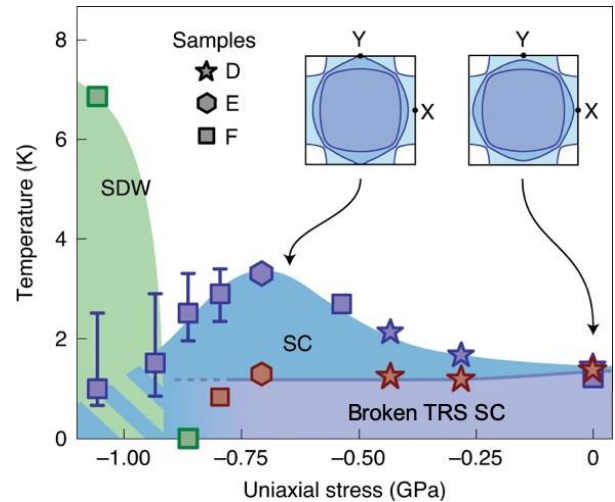

**Figure 4.** Temperature vs. stress phase diagram of SRO214. The candidates for even–parity superconducting order parameters that break TRS are $d + id$, $s + id$ and $d + ig$. The insets illustrate the Fermi surfaces of SRO214 at zero stress and stress where the maximum increase in the $T_c$ is observed. Here, SDW stands for spin density wave. (Adopted from [47] copyright NPG2021).

The current–phase relationship of SRO214 Josephson junctions should be explored to investigate TRS. For a Josephson junction of a conventional and an unconventional superconductor, the inversion of $I_c$ with the current polarity and magnetic field can provide evidence for broken TRS; three different inversion symmetries are presented in Figure 5a. Kashiwaya et al. [72] investigated planar and corner SRO214/Nb junctions and found that TRS is invariant, which is consistent with helical superconductivity (Figure 5b,c). Furthermore, Uchida et al. [73] investigated point contact weak links with a cross-sectional area of 0.2 µm$^2$ using SRO214 thin films. The critical–current density $J_c$ was of the order of $10^5$ A/cm$^2$, oscillating with a magnetic field along the *c*-axis is also indicating invariant TRS. Moreover, for these junctions, a temperature–dependent $I_c$ may indicate that the order parameter is *s*-wave or chiral *d*–wave [73]. The $J_c$ is much lower than that observed in the junctions based on SRO214 single crystals, e.g., Saitoh et al. [34] reported $10^7$ A/cm$^2$. This may relate to the poor interface quality that complicates the current-phase relation, making it difficult to draw a conclusive picture of the order parameter.

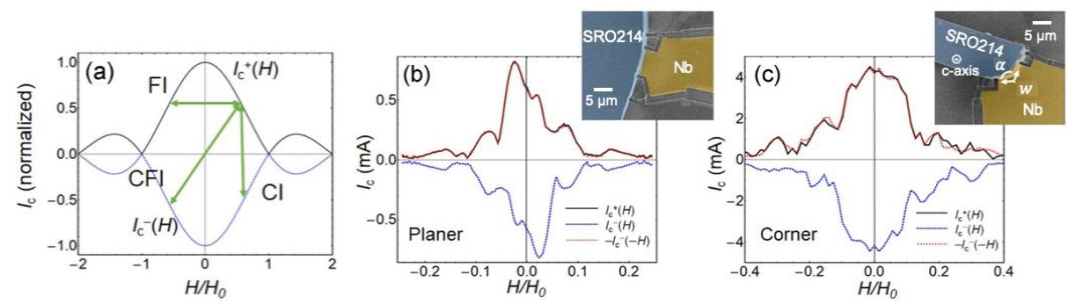

**Figure 5.** (**a**) Critical current vs. normalised magnetic field ($I_c(H/H_\circ)$) showing three types of symmetries; including, current inversion (CI), magnetic field inversion (FI), and current and magnetic field inversion (CFI). (**b**,**c**) $I_c(H)$ measurements for planner ($T$ = 0.31 K, $w$ = 7.1 µm) and corner ($T$ = 0.32 K, $w$ = 10 (5 + 5) µm, $\alpha$ = $2/3\pi$) junctions. Consistency between $I_c^+(H)$ (black) and $-I_c^-(-H)$ (red) curves indicates that the critical current in both junctions exhibits CFI symmetry. It supports the time-reversal invariance of superconductivity in SRO214. False-colour scanning ion micrographs of the junctions are shown in the insets. (Adopted from [72] copyright APS2019)

## 5. Dynamics of Chiral Domains Walls

Instabilities of $I_c$ in SRO214 junctions could be chiral–domain wall dynamics. Kidwingira et al. [31] investigated SRO214/Nb in–plane junctions and observed anomalous hysteresis in $I_c(H)$ (Figure 6a) and sharp switching in $I_c$ vs. out–of–plane magnetic field, as shown in Figure 6b,c. Close to $T_c$, a junction oscillates between normal and superconducting states generating telegraphic–like noise (Figure 6d), suggesting that magnetic-field loops encourage domain motion. Note that the junctions were fabricated using a pure SRO214 single crystal and Nb electrodes were sputtered onto the *ac*-plane, as shown in Figure 6e. The results of these experiments could be simulated using different configurations of chiral–domain walls, as illustrated in Figure 6f,g.

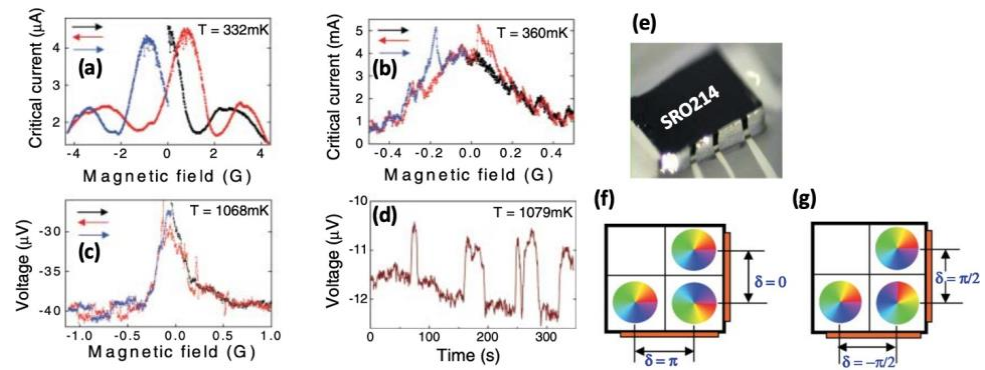

**Figure 6.** $I_c(H)$ of Pb/SRO214 junctions with a pure single crystal SRO214. (**a**) At low temperatures (332 mK), $I_c(H)$ shows a hysteresis and (**b**) sharp $I_c$ transitions. (**c**) Close to $T_c$, the hysteretic behaviour is decreased while switching of $I_c$ is still observed. (**d**) Telegraphic–like noise in the voltage at fixed bias currents above $I_c$ at zero applied magnetic field. (**e**) A scanning electron micrograph of the junction. (**f**,**g**) Configurations of chiral domains utilised for simulations; explained in the text. (Adopted from [31] copyright AAAS2006)

The temperature dependence of $I_c$ is nonmonotonic for a Josephson junction with two superconductors of different superconducting symmetries. To identify the symmetry of SRO214, Jin et al. [74] investigated the temperature dependence of $I_c$ of Pb/SRO214/Pb Josephson junctions. The junctions are fabricated using a shadow mask and deposition of two *s*-wave superconductor Pb electrodes at a polished *ab*-surface of an SRO214–Ru eutectic single crystal. A Ru inclusion provides a metallic contact between Pb and Ru; otherwise, Pd does not have a good electronic contact to SRO214; a schematic junction is shown in the inset of Figure 7b. In these junctions, at $T_{c-SRO214} < T < T_{c-Pb}$ the junction is in the S/N/S configuration as SRO214 is in the normal state. In this regime, $I_c$ increases monotonically with decreasing temperature. However, at $T = T_{c\ SRO214}$ the $I_c$ sharply reduces to zero and increases again with further decrease in the temperature, as shown in Figure 7a,b. Similar suppression of the $I_c$ has been reported in Pb/Ru/SRO214 junctions [75,76] (see Figure 7c). Theoretically, Yamashiro et al. [77] demonstrated that such a nonmonotonic $I_c$ is due to the interference of order parameters of a *p*–wave spin–triplet and an *s*–wave spin–single superconductor.

Note that in the above–mentioned junctions [74–76], a large number of Ru inclusions of different aspect ratios are involved. This means that there is a number of parallel junctions available for electronic transport. Therefore, the geometry of the junctions is not well defined, and the effect of inhomogeneous interfaces cannot be ignored. To avoid such geometrical effects, Anwar et al. [35,36] developed Nb/Ru/SRO214 junctions by depositing Nb (*s*–wave superconductor) only on a single Ru inclusion. Such junctions are topological as the electronic transport is dependent on the relative phase difference between two superconductors. A schematic illustration of a junction is shown in the inset of Figure 7g. The transport properties were investigated vs. temperature, magnetic fields, and junction size. Telegraphic noise emerged below $T_c$ in the 1.5–K–phase. For the 3-K–phase ($T_c < T < T_{e\ 3K}$), the junctions do not show instability in $I_c$ (see Figure 7g,h). It has also been reported that the junctions with smaller areas ($\approx 10\ \mu m^2$) are stable, indicating that the size of a domain is in the range of 5 μm [36]. The switching in $I_c$ is dependent on the sweep direction in the magnetic field [78], as shown in Figure 7d–f. The Ru-inclusions can also provide pinning sites for chiral domain walls [36].

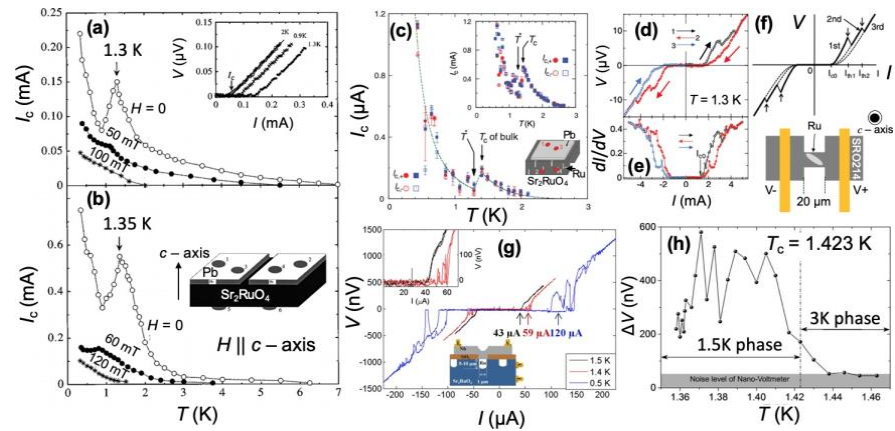

**Figure 7.** Critical current behaviour of Josephson junctions with SRO214–Ru eutectic single crystals. (**a**,**b**) Temperature–dependent critical current ($I_c(T)$) of the Pb/SRO214/Pb Josephson junction, as shown in the inset. A sharp $I_c$ suppression at bulk $T_c$ is observed, which is suppressed with the applied magnetic fields. (Adopted from [74] copyright APS 1999) (**c**) $I_c(T)$ of a Pb/Ru/SRO214 junction indicating similar $I_c$ suppression. The inset shows that $I_c(T)$ is stable at $T > T_{c-bulk}$, but it is unstable for $T < T_{c-bulk}$. (Adopted from [75] copyright APS 2011) **d**,**e**) Current–voltage (*IV*) curves and differential conductance of a junction prepared using FIB and keeping a single Ru inclusion between two SRO214 bulk parts. Close to the $T_c$ (1.3 K), sweep sensitive switching in current and $dV/dI$ are observed. (**f**) A schematic of $I_c$ switching and configuration of the junctions. (Adopted from [78] copyright APS 2008). (**g**) IV curves at different temperatures of a Nb/Ru/SRO214 topological junction based on a single Ru inclusion, as shown in the inset. The $I_c$ is switching between superconducting and normal states. (Adopted from [35] copyright NPG 2013) (**h**) Temperature–dependent switching effect that is totally suppressed in the 3–K–phase. (Adopted from [36] copyright APS 2017).

In an unstable state of SRO214 topological junctions, current-voltage $I(V)$ curves are anisotropic in current. In addition to a suppression of $I_c$, Nakamura et al. [75,76] observed anisotropic $I(V)$ curves ($I_{c+} \neq I_{c-}$) at $T < T_{c-bulk}$ and instabilities in $|I_c|$ below $T_c$. These instabilities may arise because of the topological behaviour (phase winding) of an interface between an *s*-wave and a chiral *p*-wave superconductor. For $T > T_{c\ bulk}$, the phase winding number $N = 0$ as a phase of the *s*-wave superconductor (proximitised Ru inclusion) matches with the *p*-wave superconductor (see Figure 8a). At lower temperatures, the chiral *p*–wave superconductor (here, it is SRO214) achieves bulk superconductivity and with phase winding $N = 1$. Just at the bulk transition, a significantly large phase difference $\delta\varphi$ suppresses the $I_c$ (see Figure 8b,c). In principle, this topological effect could also be justified with the chiral *d*-wave superconducting order parameter.

The telegraphic noise in voltage (switching between superconducting and normal states) can be explained with the motion of chiral–domain walls, as shown in Figure 8d,e. A chiral–domain wall can be excited with external stimuli, such as temperature [35,36], applied magnetic field [31], excitation current [35] and geometry of the junction [36]. The main issue of the nature of the symmetry of the order parameter cannot be identified only based on such telegraphic-like noise because such noise can equally be explained with the helical *p*–wave spin–triplet or chiral *d*–wave spin-single symmetries. However, interestingly, these features of telegraphic–like noise and instabilities in the $I_c$ are strongly indicating the domain formation but only in the bulk SRO214 (1.5 K-phase). Comparing it with the recent observation of broken TRS in SRO214 may suggest that SRO214 exhibits chiral superconductivity of either *d*–wave or *p*-wave.

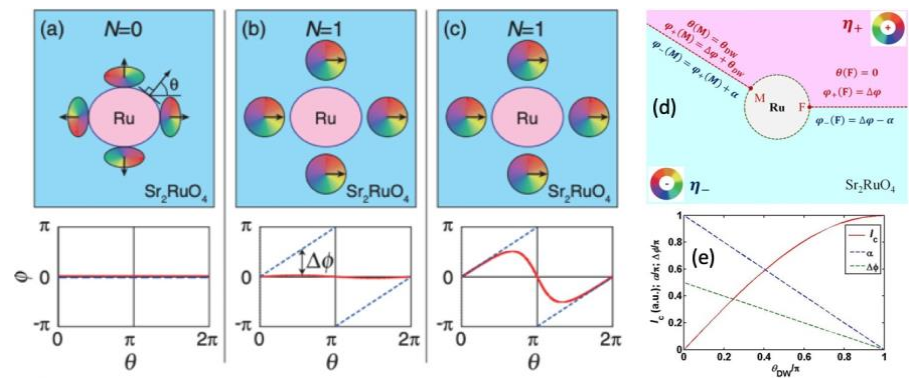

**Figure 8.** (**a**–**c**) Schematic illustrations of the superconducting order parameter at the SRO214/Ru interface. Upper panels present superconducting phase $\varphi$ as a function of direction ($\theta$: normal to the surface) at each spatial position; arrows indicate the direction of momentum for $\varphi = 0$. Lower panels illustrate the directional dependence of $\varphi(\theta)$ at the SRO214/Ru interface in equilibrium state (without external current). The solid (dotted) lines represent the $\varphi$ in Ru ($\varphi$ in SRO214). (**a**) At $T_{c-bulk} < T < T_{3K-phase}$, the winding number $N = 0$: $\varphi$ in Ru and SRO214 match. (**b**) At $T_{c-bulk} \lesssim T$ phase winding $N = \pm 1$, which is increasing the phase difference $\delta\varphi$. (**c**) At $T < T_{c-bulk}$, the interfacial energy is increasing, which deforms $\delta\varphi$. (Adopted from [75] copyright APS 2011) (**d**) Schematic of two chiral-domain walls in SRO214 intersecting a circular Ru inclusion. The domain wall "F" is fixed at angle $\theta = 0$, while domain wall "M" is movable with $\theta = \theta_{DW}$, where $\alpha$ is the phase difference across a domain wall. (**e**) Calculated maximum $I_c$, $\alpha$ and $\delta\varphi$ depending on $\theta = \theta_{DW}$. (Adopted from [35] copyright NPG 2013)

## 6. Proximity Structures

Electronic transparency between two different oxide materials can be challenging to achieve. Electronic transparency between a normal metal and an oxide is defined by the Fermi surface mismatch and electronegativity of the metal. SRO214 develops a poor electronic contact with most metals, particularly when deposited on the *ab*–surface [33,35,75,76]. One reason is atomic reconstructions that result in RuO$_6$ octahedra rotation of $\approx 3°$ around the *c*–axis [79]. This octahedral rotation suppresses the superconductivity. Note that the rotation happened only in a few top layers that laterally displaces the oxygen atoms in the RuO$_6$ plane up to 0.3 Å compared with the position in the bulk. This atomic reconstruction is one of the main hurdles in developing superconducting interfaces between SRO214 and other materials to investigate the proximity effect. Moreover, theoretical calculations show that a rotation of more than 6.5° causes band narrowing that increases the density–of–states at the Fermi–level and stabilises the ferromagnetic order with a magnetisation of the order 0.1 $\mu_B$ [79]. So far, such a magnetic transition has not been reported.

Various research groups have investigated electronic transport in SRO214 junctions fabricated by sputtering *s*–wave superconductors or normal metals on an SRO214 single crystals both on the *ab*-surface (*c*-axis oriented junction: current flow along the *c*-axis) and *ac*–surface (in-plane junctions: current flows along the *ab*-plane) [31,32]. The in–plane junctions are investigated because of two reasons: (1) a relatively good interface the *ac*-surface and a normal metal can be developed, (2) by considering a quasi–two–dimensional superconductor in SRO214, mainly in–plane transport properties are important [32].

An atomically smooth and electronically transparent interface can be achieved by epitaxially growing a metallic oxide on SRO214. To probe in–plane transport properties, a thin film must be grown on the *ac*–surface, but this is challenging, since SRO214 is a layered material with a 6.4 Å lattice parameter. However, there are many candidate materials (e.g., CaRuO$_3$, SrRuO$_3$, La$_{1-x}$Ca$_x$MnO$_3$, La$_{1-x}$Sr$_x$MnO$_3$) that can be deposited epitaxially on the *ab*–surface as the *a*–axis mismatch is significantly lower. For example, SRO113 has a *a*–axis mismatch with SRO214 of about 1.5%, exhibiting the same type of RuO$_6$ octahedral. Anwar et al. [80] reported epitaxial growth of SRO113 thin films on a leaved *ab*–surface of an SRO214 single crystal using pulsed laser deposition (PLD). Furthermore, they observed that the SRO113/SRO214 interface is atomically smooth and a high conductor as well. This indicates that deposition of SRO113 may suppress the $RuO_6$ octahedral rotation, which improves the electronic state of the interface.

Atomic reconstruction occurs even after cleaving at low temperatures (70 K) and in high vacuum. At the film interface, $RuO_6$ elongated about 0.1 Å. Siwakoti et al. [81] used LEEM to investigate the $RuO_6$ octahedral rotation in SRO214 as a substrate. They observe (1 × 1) integer spots and also ($\sqrt{2} - \sqrt{2}$)R45 fractional spots on the *ab*–surface. Fractional spots indicate the $RuO_6$ octahedral rotation, which is still unchanged even after treating the substrate with deposition conditions, such as heating the substrate at 700 °C under $O_2$ pressure of 2 mBar. Interestingly, fractional spots disappear just after the deposition of LSMO thin films, and the $RuO_6$ octahedral has elongated about 0.1 Å along the *c*–axis. This indicates that epitaxial growth engineers the SRO113/SRO214 interface [80,82].

Anwar et al. [82] demonstrated a long–range proximity effect at an epitaxial SRO113/SRO214 interface with a coherence length of the order of 9 nm in a 15-nm thick SRO113 layer (Figure 9). This indicates that induced superconducting correlations in SRO113 are spin-triplet, as the coherence length for spin–single correlation in SRO113 is not more than 1 nm in the diffusive limit (3 nm in clean limits). Interestingly, the same group reported an anisotropic response of the proximity effect with the direction of externally applied magnetic fields [83]. Moreover, they investigated the gap feature of induced superconductivity into SRO113 using Au/STO/SRO113/SRO214 tunnel junctions [84]. $dI/dV$ versus bias voltage may indicate that the induced superconductivity in SRO113 has an anisotropic gap function such as *p*-wave or *d*-wave symmetry.

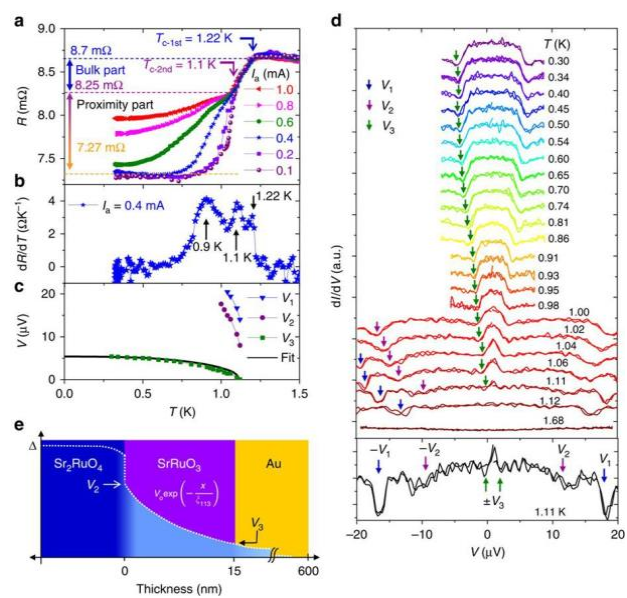

**Figure 9.** (**a**) Temperature-dependent resistance $R(T)$ of a Au/SRO113/SRO214 junction in the range of 1.5–0.3 K, measured with different applied currents (0.1–1 mA). Two obvious transitions are observed; first, the robust applied current corresponds to bulk, and the second transition corresponds to the proximity effect. (**b**) $dR/dT$ indicating various transitions in the junction. (**c**) Evolution of differential conductance $dI/dV$ with temperature. (**d**) Voltage biased $dI/dV$ at different temperatures. The curves are shifted vertically for clarity. The lower panel shows a $dI/dV$ at 1.11 K to emphasise features. (**e**) A schematic illustration of the proximity effect in the junction. (Adopted from [82] copyright NPG 2016).

Considering that SRO214 is a spin–singlet instead of spin-triplet, in that case, a magnetic inhomogeneity is an essential requirement for singlet–to–triplet conversion [4], which leads to a long-range proximity effect. This is difficult to achieve with an epitaxial SRO113/SRO214 interface, particularly when a single ferromagnetic layer is used. How-ever, broken inversion symmetry at the SRO113/SRO214 interface can produce the Rashba type SOC with an in–plane SOC effective field. In this configuration of the effective SOC field and out–of–plane magnetic anisotropy of SRO113, the interface may exhibit the spin activity required to generate spin–triplet correlations, which results in a long–range proximity effect. In the clean limit (thickness of the ferromagnetic layer $< l_e$), the even-frequency $p$–wave superconductivity may dominate over the $s$–wave odd–frequency correlations [5]. That may explain the long–range proximity effect observed in Reference [82].

## 7. Candidate Materials for Unconventional Superconducting Junctions

Before concluding our discussion, we briefly overview possible candidate materials for spin–triplet superconductivity and their suitability for developing Josephson junctions, tunnel junctions, and SQUIDs, to investigate the superconductivity and potential applications in quantum information and superconducting spintronic technologies.

The coexistence of ferromagnetism and superconductivity has been reported in Uranium compounds, including $UPt_3$, $UTe_2$, UGe, URhGe, and UCoGe [85]. In these compounds, higher values of upper critical field ($H_{c2}$) and re–entrant of superconductivity in some of these compounds at higher fields indicate the existence of spin–triplet superconductivity [86]. For the $UTe_2$ superconductor at critical temperature ($T_c$) of 1.6 K nuclear magnetic resonance (NMR), the Knight shift is unchanged in the superconducting state relative to the normal state [18]. Furthermore, Kerr rotation measurements reveal the internal field below $T_c$, indicating that TRS is broken. These findings support the spin-triplet superconductivity in $UTe_2$ [18]. Since these are Uranium compounds, it is restricted to utilising such materials independently to develop the junctions. Still, few groups utilised $UPt_3$ to study the proximity effect of a heavy fermion superconductor that indicates very low interface transparencies due to fermi surface mismatch at the interface [87].

There is another list of heavy fermion materials, such as Cerium compounds ($Ce_nM_mIn_3n + 2$) and Plutonium compounds ($PuMX_5$) for which unconventional superconductivity has been reported [88–90]. The symmetry of the superconducting order parameter of some of these compounds is still controversial. In addition to the single crystals, thin films of heavy fermion superconductors can be grown using molecular beam epitaxy (MBE) or sputtering as well [91–93], which can be utilised to develop spin-triplet junctions to investigate the superconducting properties and superconducting spin transport. However, the higher effective mass of heavy fermions suppresses the interface transparency and makes it painstaking to develop sufficiently good quality junctions for fundamental and applied research work [87].

The crystal structure of non–centrosymmetric superconductors (NCS) lacks inversion symmetry and such materials tend to have strong SOC [94]. Parity in NCS materials is therefore no longer a good quantum number, which leads to a mixture of even–parity spin-singlet and odd–parity spin–triplet states [95]. NCS materials are promising candidates for spin–triplet superconducting junctions. However, their mixed parity state complicates the junction properties.

A topological insulator (TI) has an open gap in the bulk and non–trivial gapless edge or surface states that depend on SOC [96]. Similarly, a superconductor can also hold non–trivial topological gapless states with a superconducting gap in the bulk [97,98]. Interestingly, a three–dimensional topological insulator $Bi_2Se_3$ shows superconductivity under pressure [99], which is expected to be a topological superconductor. Therefore, it is also expected that doped $Bi_2Se_3$ via intercalation of Cu, Sr or Nb may exhibit a topological spin–triplet superconducting order parameter [100]. However, it is observed that the TRS is invariant for $Cu_xBi_2Se_3$ and it is broken for $Sr_{0.1}Bi_2Se_3$ [101]. This may indicate that $Cu_xBi_2Se_3$ is not in a triplet state, and $Sr_{0.1}Bi_2Se_3$ might exhibit a singlet triplet mixed state or a purely topological spin–triplet superconductor.

## 8. Summary

We have reviewed electronic transport in superconducting junctions with pure and eutectic SRO214 single crystals. The vast majority of junctions exhibit anomalous instabilities in $I_c$, and telegraphic-like noise below $T_c$—in the 1.5 K-phase. This indicates that the bulk superconductivity of SRO214 exhibits a dynamical order-parameter. In the 3–K–phase, the $I_c$ remains stable with a monotonic temperature dependence [35,36,74–76]. These reports may indicate that both the 1.5–K– and 3–K–phase have distinct superconducting order parameters, namely dynamic and static, respectively. The broken TRS and the dynamic behaviour of the junctions based on SRO214 indicate chiral domains. Such domains can emerge due to chiral superconductivity in SRO214 in the $p$–wave spin–triplet or $d$–wave spin–singlet state. Unfortunately, the spin degree of freedom of the order parameter cannot be identified through existing transport studies of the junctions; however, recent observations of reduced electronic spin susceptibility may only support the spin-single scenario. A possible straightforward pathway can be investigations of density–of–states of a $c$–axis-oriented tunnel junction that may exhibit unique zero bias peaks. Moreover, spin sensitive transport properties of a non–locale junction may identify the spin–state of the order parameter.

Thus far, junctions based on SRO214 are mainly prepared using cleaved or eutectic single crystals that may not provide a well–defined orientation of the junctions to develop a clearer picture of the superconducting order parameter of SRO214, and well–controlled superconducting junctions are required. For this, superconducting thin films in SRO214 can play a crucial role. Thin films will enable the creation of heterostructure thin–film stacks with multiple different oxide materials for studies of proximity junctions. Recent progress in the growth of SRO214 thin films and epitaxial growth of various oxides on SRO214 single crystals opens up new opportunities to develop SRO214–based junctions.

**Author Contributions:** M.S.A and J.W.A.R. contributed equally to the review article. All authorshave read and agreed to the published version of the manuscript.

**Funding:** This work was supported by the JSPS-EPSRC core-to-core programme "Oxide-Superspin (OSS)" (EP/P026311/1) and the EPSRC Programme Grant (EP/N017242/1).

**Institutional Review Board Statement:** Not applicable.

**Informed Consent Statement:** Not applicable.

**Data Availability Statement:** Data are contained within the article or cited otherwise.

**Acknowledgments:** We are thankful for valuable discussions with D. Manske, Y. Tanaka, M. Cuoco, Y. Maeno, T. W. Noh and S. Yonezawa.

**Conflicts of Interest:** The authors declare no conflict of interest.